SUBJECT AREAS

Correspondence and requests for materials should be addressed to jianwangphysics@pku.edu.cn (Jian Wang) and kehe@iphy.ac.cn (Ke He)

# Demonstration of surface transport in a hybrid $Bi_2Se_3$/$Bi_2Te_3$ heterostructure

Yanfei Zhao[1#], Cui-Zu Chang[2,3#], Ying Jiang[4], Ashley DaSilva[5], Yi Sun[1], Huichao Wang[1], Ying Xing[1], Yong Wang[4], Ke He[2,*], Xucun Ma[2], Qi-Kun Xue[2,3], and Jian Wang[1,*]

[1]International Center for Quantum Materials, School of Physics, Peking University, Beijing 100871, People's Republic of China

[2]Institute of Physics, Chinese Academy of Sciences, Beijing 100190, China

[3]State Key Laboratory of Low-Dimensional Quantum Physics, Department of Physics, Tsinghua University, Beijing 100084, China

[4]Center of Electron Microscopy, State Key Laboratory of Silicon Materials, Department of Materials Science and Engineering, Zhejiang University, Hangzhou, 310027, China

[5]Department of Physics, University of Texas at Austin, Austin, Texas 78712-1081, USA

[#]Authors equally contributed to this work.

## Abstract

In spite of much work on topological insulators (TIs), systematic experiments for TI/TI heterostructures remain absent. We grow a high quality heterostructure containing single quintuple layer (QL) of $Bi_2Se_3$ on 19 QLs of $Bi_2Te_3$ and compare its transport properties with 20 QLs $Bi_2Se_3$ and 20 QLs $Bi_2Te_3$. All three films are grown on insulating sapphire (0001) substrates by molecular beam epitaxy (MBE). *In situ* angle-resolved photoemission spectroscopy (ARPES) provides direct evidence that the surface state of 1 QL $Bi_2Se_3$ / 19 QLs $Bi_2Te_3$ heterostructure is similar to the surface state of the 20 QLs $Bi_2Se_3$ and different with that of the 20 QLs $Bi_2Te_3$. In *ex situ* transport measurements, the observed linear magnetoresistance (MR) and weak antilocalization (WAL) of the hybrid heterostructure are similar to that of the pure $Bi_2Se_3$ film and not the $Bi_2Te_3$ film.

This suggests that the single $Bi_2Se_3$ layer on top of 19 QLs $Bi_2Te_3$ dominates its transport properties.

**Introduction**

Three dimensional (3D) TIs are band insulators with gapless linear energy dispersion surface states[1-2]. $Bi_2Se_3$, $Bi_2Te_3$ and $Sb_2Te_3$ are confirmed as typical 3D TIs due to their simple single surface Dirac cone and relatively large bulk energy gap[3-4]. These materials provide the possibility for the observation of novel phenomena, including Majorana fermions[5], magneto-electric effect[6-7], and quantum anomalous Hall effect[8-9], as well as the potential applications in quantum computation[10]. We know that semiconductor heterostructure consists of materials with different band structures and exhibits interesting quantum behaviors, such as quantum Hall effect[11] and fractional quantum Hall effect in $GaAs/Al_xGa_{1-x}As$ heterostructures[12]. Recently, normal band insulator-3D TI or semimetal-3D TI heterostructures, such as $Sb_2Se_3$-$Bi_2Se_3$[13] and Bi-$Bi_2Se_3$ ($Bi_2Te_3$) heterostrctures[14-15], have attracted much interest since the interface offers a new platform to study topological phase. However, up to now few experiments have been carried on the heterostructure of two different TIs. According to TI band theory, the existence of gapless surface state is due to the topological invariant change in the interface of trivial and non-trivial insulators[1]. As the ARPES measurements reported before[16], when the thickness of $Bi_2Se_3$ is below 6 QLs, the surface state opens a gap at the Dirac Point (DP) due to the coupling between the top and bottom surfaces. Therefore, when 1QL $Bi_2Se_3$ is grown on a trivial insulator, the Dirac cone of the surface state of $Bi_2Se_3$ cannot survive. Then, a natural question is how about 1 QL $Bi_2Se_3$ film on another type of TI. On the other hand, when we just replace 1 nm thick surface layer of one TI by a different TI, what new property will come into being?

In this work, we epitaxially grow 1 QL $Bi_2Se_3$ on 19 QLs $Bi_2Te_3$ and form a 1 QL $Bi_2Se_3$ / 19 QLs $Bi_2Te_3$ heterostructure[17]. *In situ* ARPES results show that the surface state of the heterostructure 1 QL $Bi_2Se_3$ / 19 QLs $Bi_2Te_3$ is similar to the surface state of high electron doped $Bi_2Se_3$ film. Further *ex situ* measurements demonstrate that the transport properties of $Bi_2Se_3$ / $Bi_2Te_3$ heterostructure, such as linear MR and WAL of our $Bi_2Se_3$ / $Bi_2Te_3$ heterostructure behave totally different from that of pure $Bi_2Te_3$, but similar to pure $Bi_2Se_3$. The experiments show that the 1 QL $Bi_2Se_3$ top layer dominates the transport properties of the heterostructure and clearly demonstrates the effect of the surface state on the transport of a TI.

## Results

The TI thin films studied here are grown on the sapphire (0001) substrate by MBE[17]. Transport measurements are performed in a commercial physical property measurement system (PPMS-16T). The ARPES band dispersions and corresponding momentum distribution curves (MDCs) of 20 QLs $Bi_2Se_3$, 20 QLs $Bi_2Te_3$ and 1 QL $Bi_2Se_3$/19 QLs $Bi_2Te_3$ are shown in **Fig. 1a** to **1f**, respectively. As shown in **Fig. 1a** and **1d**, the DP of the 20 QLs $Bi_2Se_3$ lies in the bulk gap and is located at 0.17 eV below the Fermi level ($E_F$). On the other hand, the ARPES band map of 20 QLs $Bi_2Te_3$ is different. The DP is buried in the bulk valance band and the $E_F$ lies near the bulk conduction band due to the n-type carriers caused by the Te vacancies (**Fig. 1b** and **1e**). As reported before, when the thickness of $Bi_2Se_3$ is below 6 QLs, the surface state opens a gap at the DP due to the coupling between the top and bottom surfaces[16]. However, when 1 QL $Bi_2Se_3$ is epitaxially grown on 19 QLs $Bi_2Te_3$, a single gapless Dirac cone is observed in the bulk gap. The surface band dispersion near the DP, especially the position of DP is almost same with the surface state of electron doped $Bi_2Se_3$ films (**Fig. 1c** and **1f**). The significant change in the surface band dispersion from that of $Bi_2Te_3$ to that of $Bi_2Se_3$, instead of co-existing of them, indicates that the 1 QL of $Bi_2Se_3$ basically covers the whole surface of $Bi_2Te_3$. The electron doping of 1QL $Bi_2Se_3$ is likely to be induced by the underneath $Bi_2Te_3$ which has higher electron doping level than that of $Bi_2Se_3$. Thus, the ultrathin 1 QL $Bi_2Se_3$ changes the surface state of the top surface of 19 QLs $Bi_2Te_3$ film and makes the 1 QL $Bi_2Se_3$ / 19 QLs $Bi_2Te_3$ heterostructure exhibit $Bi_2Se_3$ surface state characteristic. This observation is reasonable since surface states are localized near the surface region of a material, and therefore the surface band dispersion is mainly determined by the environment around surface.

1 QL $Bi_2Se_3$ / 19 QLs $Bi_2Te_3$ heterostructure film was examined by a FEI TITAN Cs-corrected STEM operating at 200 kV. **Figure 2a** shows a high angle annular dark field (HAADF) image of the heterostructure film with amorphous Te capping layer and sapphire substrate. **Figure 2b** and **2c** show the atomic layer-by-layer HAADF images which manifest a high-quality single crystal nature of 1 QL $Bi_2Se_3$ / 19 QLs $Bi_2Te_3$ heterostructure film. According to the Z contrast, we infer the van der Waals gap indicated by the green dashed line is the interface between $Bi_2Se_3$ and $Bi_2Te_3$. Besides, the interfaces of Te capping layer / TI film and TI film / sapphire are clearly defined due to the different atomic structures, as marked by red and yellow dashed lines.

## Discussion

To further study the properties of 1 QL Bi$_2$Se$_3$ / 19 QLs Bi$_2$Te$_3$ heterostructure, we performed transport measurements on 20 QLs Bi$_2$Se$_3$, 20 QLs Bi$_2$Te$_3$ and 1QL Bi$_2$Se$_3$ / 19QLs Bi$_2$Te$_3$ heterostructure film for comparison. **Figure 3a** is a schematic diagram of our transport measurement structure. Standard four probe method was used in all the measurements. We placed two indium current electrodes (*I+* and *I-*) on both ends and across the entire width of the film strip with a size of 0.3mm×2mm, so that the current can homogeneously go through the TI film in longitudinal direction. Then, two indium electrodes with the diameter of 0.3 mm were pressed on the film as voltage probes. In this paper, all the transport results are repeatable in different MBE-grown samples. Hall measurement results are shown in Fig. S1 in Supplementary Information. **Figure 3b** shows the temperature dependence of two dimensional (2D) sheet resistance ($R_{sq}$) for 20 QLs Bi$_2$Se$_3$ and 20 QLs Bi$_2$Te$_3$. In pure 20 QLs Bi$_2$Se$_3$ and 20 QLs Bi$_2$Te$_3$ films, with decreasing temperature (*T*), $R_{sq}$ displays metallic behavior at high *T* region and becomes weakly insulating at low *T* regime. In **Fig. 3c**, the 1 QL Bi$_2$Se$_3$ /19 QLs Bi$_2$Te$_3$ film shows the similar metallic behavior with the pure Bi$_2$Te$_3$ film at high *T*. However, it displays the stronger resistance upturn than the 20 QLs Bi$_2$Te$_3$ film at low *T*. Under a perpendicular magnetic field, the resistance upturn still exists for all three samples. It is reminiscent of the electron-electron interaction in TI films[18]. However, the slopes of the normalized resistance ($R/R_{min}$) – ln*T* for 20 QLs Bi$_2$Se$_3$ and 20 QLs Bi$_2$Te$_3$ are different. For 20 QLs Bi$_2$Se$_3$ (**Fig. 3d**), the slope of $R/R_{min}$ – ln*T* shows obvious magnetic field-dependence. The dashed lines are eye-guides to show that $R/R_{min}$ increases logarithmically with *T* in low *T* regime. With increasing magnetic field, the resistance upturn is decreased and almost suppressed at 100 kOe. Nevertheless, as shown in **Fig. 3e**, the slope of $R/R_{min}$ – ln*T* of 20 QLs Bi$_2$Te$_3$ shows weak dependence of the external magnetic field and the resistance upturn is enhanced by the field. **Figure 3f** shows the normalized resistance versus *T* at different magnetic fields for 1 QL Bi$_2$Se$_3$ / 19 QLs Bi$_2$Te$_3$ heterostructure film. Compared with 20QLs Bi$_2$Te$_3$, the heterostructure shows more obvious upturn behavior which is nearly independent of magnetic field when the field is larger than 10 kOe. The slope of the resistance upturn is defined by $\kappa = (\pi h/e^2)d\sigma_{xx}/d\ln T$ [19]. As shown in the inset, $\kappa$ increases with increasing the magnetic field and becomes a constant when the magnetic field ranges from 10 kOe to 150 kOe. The electron-electron interaction causes a logarithmic increase of resistance while WAL induces a logarithmic decrease of resistance[18,20]. The two effects coexist and compete with each other. Thus, at zero field both WAL and electron-electron interaction cause ln*T* dependence which results a small $\kappa$. At higher magnetic field, the WAL contribution is suppressed and $\kappa$ becomes a constant due to the strong electron-electron interaction (The complete data are shown in Fig. S2 in Supplementary Information).

**Figure 3g** and **3h** show the normalized MR ($R$(H)/$R$(0)) of 20 QLs $Bi_2Se_3$ and 20 QLs $Bi_2Te_3$ thin films as a function of magnetic field perpendicular to thin films at different $T$. A linear and non-saturating MR is observed in the field up to 150 kOe for 20 QLs $Bi_2Se_3$ in **Fig. 3g**, which even maintains at the temperature above 50 K. On the other hand, as shown in **Fig. 3h**, the 20 QLs $Bi_2Te_3$ film displays a clearly nonlinear MR behavior and the value of normalized MR ($R$(H)/$R$(0)) is 2.18 at 2 K and 150 kOe, which is much larger than that of the 20 QLs $Bi_2Se_3$ film (1.3). The nonlinear MR is a typical behavior found in MBE-grown $Bi_2Te_3$ thin films[21-22], though the linear MR behavior was ever observed in bulk $Bi_2Te_3$[23-25]. The normalized MR of the 1 QL $Bi_2Se_3$ / 19 QLs $Bi_2Te_3$ film shows a linear, non-saturated and weakly temperature-dependent behavior across a wide range of magnetic field from 10 kOe to 150 kOe (**Fig. 3i**). According to Abrikosov's model, the linear and positive MR is expected to exist in the gapless semiconductor with a linear $E$-$k$ dispersion[26]. In addition, the linear MR behavior of the 1 QL $Bi_2Se_3$ / 19 QLs $Bi_2Te_3$ film gradually decreased when the sample was tilted away from the perpendicular angle and eventually became nonlinear in the parallel magnetic field. (Fig. S3 in Supplementary Information) Hence, the linear MR probably comes from the 2D gapless TI surface state[27-29]. However, the MR of 20 QLs $Bi_2Te_3$ (**Fig. 3h**) shows an obviously nonlinear behavior while 20 QLs $Bi_2Se_3$ (**Fig. 3g**) displays a linear and positive MR. Therefore, it is believed that the 1 QL $Bi_2Se_3$ on the top of heterostructure plays a significant role in the transport property. The angular dependent MR of 20 QLs $Bi_2Se_3$ and 20 QLs $Bi_2Te_3$ thin films at 2 K and 160 kOe are shown in **Fig. 3j** and **3k**. It is known that the 2D surface states of the TIs are only sensitive to the perpendicular component of the magnetic field $B\cos\theta$, where $\theta$ is the angle between magnetic field and the normal direction of the thin films (**Fig. 3j** inset). For the 20 QLs $Bi_2Se_3$ film, the observed angular dependence of MR can be well fitted by a $|\cos\theta|$ function. This suggests that the MR curve shows an obvious 2D response. Thus, the observed linear MR behavior from 20 QLs $Bi_2Se_3$ might be from the energy-momentum ($E$-$k$) linear relationship of 2D topological surface state[27]. Compared with $Bi_2Se_3$, the angular dependence of MR of 20 QLs $Bi_2Te_3$ film exhibits a little deviation with the $|\cos\theta|$ fitting curve. Moreover, the curve of 1 QL $Bi_2Se_3$ / 19 QLs $Bi_2Te_3$ heterostructure film resistance versus the rotation angle at 2 K and 150 kOe matches well with the function of $|\cos\theta|$ (**Fig. 3l**) too, implying the 2D nature of magneto-transport behavior.

TI has spin-momentum locked surface states which lead to a π Berry phase suppressing the backscattering. The absence of backscattering results in the destructive interference between the two time reversal symmetry loops, which leads to the WAL effect and the perpendicular magnetic field can destroy this quantum interference. **Figure 4** shows the WAL effect of our TI films in low field at different temperatures. With increasing $T$, the WAL effect becomes weaker and is suppressed at 50 K, as shown in **Fig. 4a** for 20 QLs $Bi_2Se_3$ and **4b** for 20 QLs $Bi_2Te_3$. Compared with two WAL effects, we find that the 20 QLs $Bi_2Se_3$ film with a sharp cusp in low field while 20 QLs $Bi_2Te_3$ film displays smaller MR dip in lower field regime. In **Fig. 4c**, the 1 QL $Bi_2Se_3$ / 19 QLs $Bi_2Te_3$ film exhibits WAL effect in the low field too. The normalized MR shows a large cusp in the perpendicular field and the behavior is weakened with increasing temperature (**Fig. 4c**), which is more like that in pure 20 QLs $Bi_2Se_3$ (**Fig. 4a**) not 20 QLs $Bi_2Te_3$ (**Fig. 4b**).

According to the Hikami-Larkin-Nagaoka (HLN) theory[30], the equation for 2D magnetoconductance (MC) is:

$$\Delta\sigma = \sigma(B) - \sigma(0) = -\frac{\alpha e^2}{2\pi^2\hbar}\left[\ln\left(\frac{\hbar}{4Bel_\phi^2}\right) - \psi\left(\frac{1}{2} + \frac{\hbar}{4Bel_\phi^2}\right)\right] \quad (1)$$

Here $\Psi(x)$ is the digamma function and $l_\phi$ is the phase coherence length. $\alpha$ is a coefficient reflecting the strength of the spin-orbital coupling and magnetic scattering. The value of $\alpha$ is 1, 0, -1/2 for the orthogonal, unitary and symplectic case, respectively. In TIs, the value of $\alpha$ indicates the type of the carriers. Considering the studies before[18,31-35], the coefficient $\alpha$ changes from -0.4 to -1.1. $\alpha = -0.5$ is owed to a single surface state and $\alpha = -1$ indicates that both top and bottom surface states contribute the transport or top surface and bulk do the contribution. In our case, $\Delta\sigma$ of 20 QLs $Bi_2Se_3$ film at 4 K in the perpendicular field fits the HLN equation quite well and yields $\alpha = -0.56$ and $l_\phi = 328$ nm (**Fig. 4d**). **Fig. 4e** shows the fitting results of 20 QLs $Bi_2Te_3$ film, whose values are $\alpha = -0.47$ and $l_\phi = 188$ nm. This suggests that one single surface channel dominates the WAL transport behavior. As shown by the red solid fitting curve in **Fig. 4f**, the fitting of 1 QL $Bi_2Se_3$ / 19 QLs $Bi_2Te_3$ film yields $\alpha = -0.2$ and $l_\phi = 182$ nm. Our extracted $-\alpha$ is smaller than 0.5, which is likely due to an enhanced electron-electron interaction of the carriers[36-37]. We have combined WAL and e-e interaction[18] in the fitting (Fig. S4 in Supplementary Information). The value of $-\alpha$ equals to 0.32 is larger than that given by HLN fitting. In addition, the value of

-$\alpha$ is still smaller than 0.5, which is indicative of increased localization effect. This could be due to a localization contribution from the bulk, for example, a bulk localization effect increased by the impurities at the interface. Besides, when the Fermi energy is close to the bulk band edge, the bulk would exhibit weak localization behavior $(\alpha > 0)$[38]. Thus, the experimentally observed "weak antilocalization" in the 1 QL $Bi_2Se_3$ / 19 QLs $Bi_2Te_3$ film may also be a combined result from both the weak antilocalization of the surface channel and weak localization of the bulk channel[38-39], which leads to a smaller -$\alpha$. We stress that these are only suggestions and further studies will need to be done to clarify the mechanism.

Because the 3D magneto-transport for bulk electrons in TI films is independent of the tilt angle of the magnetic field and the bulk electrons dominate the transport in parallel field, we subtract the 3D WAL contribution measured in the parallel field $\theta = 90^{\circ}$ from the MR measured in $\theta=0^{\circ}$, $45^{\circ}$, $60^{\circ}$, $80^{\circ}$ [31](**Fig. 4g-4i**). After removing the 3D bulk WAL effect, we can clearly observe the WAL effect caused by the 2D surface state of 20 QLs $Bi_2Se_3$, 20 QLs $Bi_2Te_3$ and 1 QL $Bi_2Se_3$ / 19 QLs $Bi_2Te_3$ at different angles in **Fig. 4g, 4h** and **4i** respectively. For 20 QLs $Bi_2Se_3$, the curves show same cusp in low field (<1kOe) which suggests that the observed WAL effect only relies on the perpendicular component of the magnetic field. In higher magnetic field, the curves deviate with each other, which is owing to the Zeeman effect[40]. Although similar WAL effect is also observed in 20 QLs $Bi_2Te_3$, the MR dip is apparent smaller than that of 20 QLs $Bi_2Se_3$. For the 1 QL $Bi_2Se_3$ / 19 QLs $Bi_2Te_3$ shown in **Fig. 4i,** it can be clearly observed that the surface state WAL behaves more like that in the 20 QLs $Bi_2Se_3$ film shown in **Fig. 4g**. After subtracting the 3D WAL effect, we can fit the MC behavior of TI films by equation (1). As shown in **Fig. 4j**, for 20 QLs $Bi_2Se_3$, $\alpha$ equals to -0.38 and $l_\phi$ is 451 nm. This is consistent with the value reported before[32-33,41]. **Fig. 4k** displays the fitting results of 20 QLs $Bi_2Te_3$ $\alpha$= -0.3 and $l_\phi$= 303 nm by subtracting the bulk effect, which is similar to previous report in $Bi_2Te_3$ films[31]. The phase coherence length becomes larger after subtracting the 3D effect, which indicates longer coherence length in surface transport channel. In **Fig. 4l,** we also fit the MC curve of 1 QL $Bi_2Se_3$ / 19 QLs $Bi_2Te_3$ from which the parallel component is removed. The value of $\alpha$ obtained from the fitting is -0.12 and the phase coherence length $l_\phi$ is 284 nm. Compared with the pure sample, the decrease of $l_\phi$ may also due to the enhanced electron-electron scattering rate in the heterostructure.

In summary, *in situ* ARPES experiments provide the direct evidence that the surface state of top surface of 1 QL $Bi_2Se_3$ / 19 QLs $Bi_2Te_3$ heterostructure is similar to the surface state of $Bi_2Se_3$. The high quality well-controlled epitaxial heterointerface was revealed by HRTEM. The transport properties of 20 QLs $Bi_2Se_3$, 20 QLs $Bi_2Te_3$ and 1 QL $Bi_2Se_3$ / 19 QLs $Bi_2Te_3$ films were studied. Both linear MR and WAL effect show that the heterostructure behaves more like $Bi_2Se_3$ even though there is only single $Bi_2Se_3$ layer grown on 19 QLs $Bi_2Te_3$ in the heterostructure. Therefore, the 1 QL $Bi_2Se_3$ in the heterostructure plays an important role in the transport property and increases the contribution of surface to some extent. Studying on this TI-TI heterostructure may provide a platform to artificially modulate the bulk and surface electronic structures of TIs respectively and pave a way for exploring potential applications in TI devices.

**Methods**

**The MBE sample growth.** The high quality 20 QLs $Bi_2Se_3$, 20 QLs $Bi_2Te_3$ and 1 QL $Bi_2Se_3$ / 19 QLs $Bi_2Te_3$ heterostructure films studied here were grown on sapphire (0001) in an ultra-high vacuum MBE-ARPES-STM combined system with a base pressure better than $2\times10^{-10}$ mbar. Before sample growth, the sapphire substrates are first outgassed at 650 °C for 90 min and then heated at 850 °C for 30 min. High-purity Bi (99.9999%) and Se (99.999%) or Bi (99.9999%) and Te (99.9999%) are evaporated from standard Knudsen cells. To reduce Se vacancies in $Bi_2Se_3$ or Te vacancies in $Bi_2Te_3$, the growth is kept in Se or Te-rich condition with the substrate temperature at 180~220 °C. When we grow the 1 QL $Bi_2Se_3$ / 19 QLs $Bi_2Te_3$ heterostructure film, 19 QLs $Bi_2Te_3$ films were firstly grown on sapphire (0001) and used as a substrate, and then 1QL $Bi_2Se_3$ film is deposited on thick 19 QLs $Bi_2Te_3$ films at the substrate temperature slightly lower (~20°C) than the growth temperature of thick 19 QLs $Bi_2Te_3$ films to reduce the possible intermixing of two different TI films.

**ARPES measurements.** The *in situ* ARPES measurements were carried out at ~150K by using a Scienta SES2002 electron energy analyzer. A Helium discharge lamp with a photon energy of hv=21.218 eV is used as the photon source. The energy and the angular resolutions were 15 meV and 0.2°, respectively. All the spectra shown in the paper are taken along the Γ-K direction. To avoid sample charging during ARPES measurements due to the insulating sapphire (0001) substrate, a 300-nm-thick titanium film is deposited at both ends of the substrate, which is connected to the sample holder. Once a continuous film is formed, the TI films are grounded through these contacts.

**Acknowledgments**

We acknowledge Moses H. W. Chan, Jainendra Jain, Nitin Samarth, Haizhou Lu, Yongqing Li, Zhong Fang, Haiwen Liu, Xinzheng Li, Junren Shi, Xincheng Xie, Ryuichi Shindou and Qingfeng Sun for fruitful discussions. This work was financially supported by National Basic Research Program of China (Grant Nos. 2013CB934600 & 2012CB921300), the National Natural Science Foundation of China (Nos. 11222434 & 11174007), and China Postdoctoral Science Foundation (No. 2011M500180 & No. 2012T50012).

**Author contribution statement**

J.W. and K.H. conceived and designed the study. Y.Z., Y.S., H.W., Y.X. and J.W. carried on transport measurement. C.C. did MBE growth and ARPES experiment. K.H., X.M. and Q.X. supervised the MBE growth and ARPES experiment. Y.J. and Y.W. made TEM study. Y.Z., C.C., A.D., Y.W. and J. W. analyzed the data. Y.Z. and J.W. wrote the manuscript.

**Additional information**

**Competing financial interests:** The authors declare no competing financial interests.

**Figure captions**

**Figure 1. ARPES results on 20 QLs $Bi_2Se_3$, 20 QLs $Bi_2Te_3$ and 1 QL $Bi_2Se_3$ / 19 QLs $Bi_2Te_3$ heterostructure thin films.** (a) 20 QLs $Bi_2Se_3$ (b) 20 QLs $Bi_2Te_3$ (c) 1 QL $Bi_2Se_3$ / 19 QLs $Bi_2Te_3$ heterostructure thin film measured along the K-Γ-K direction, the yellow line indicates the Fermi level, and the red and blue lines

indicate the surface state. The lower figures (d) to (f) are the corresponding momentum distribution curves, the blue lines indicates the Fermi level.

**Figure 2. STEM- HAADF images of 1 QL Bi$_2$Se$_3$ / 19 QLs Bi$_2$Te$_3$ film on sapphire (0001).** The dashed lines represent different interfaces (Red: Amorphous Te capping layer / TI film; Green: 1QL Bi$_2$Se$_3$ / 19QLs Bi$_2$Te$_3$; Yellow: TI film / sapphire substrate). Scale bars are 5 nm (a), 1 nm (b), 1 nm (c).

**Figure 3. Transport measurements on 20 QLs Bi$_2$Se$_3$, 20 QLs Bi$_2$Te$_3$ and 1 QL Bi$_2$Se$_3$ / 19 QLs Bi$_2$Te$_3$ heterosturcture films.** (a) Schematic structure of the TI films for transport measurements. The thickness is not to scale. (b) 2D sheet resistance ($R_{sq}$) vs temperature of 20 QLs Bi$_2$Se$_3$ and 20 QLs Bi$_2$Te$_3$ from room temperature to 2 K. (c) 2D sheet resistance ($R_{sq}$) as a function of temperature for 1 QL Bi$_2$Se$_3$ / 19 QLs Bi$_2$Te$_3$ heterostructure film. (d) - (f) The normalized resistance ($R/R_{min}$) of 20 QLs Bi$_2$Se$_3$, 20 QLs Bi$_2$Te$_3$ and 1 QL Bi$_2$Se$_3$ / 19 QLs Bi$_2$Te$_3$ as a function of temperature at different magnetic field, indicate that a logarithmic increase with decreasing $T$ (the dashed lines are guides to the eyes). In the upper inset of (f), the slope defined as $\kappa = (\pi h/e^2)d\sigma_{xx}/d\ln T$ is shown as a function of magnetic field. (g) - (i) The normalized MR ($R(H)/R(0)$) vs magnetic field of 20 QLs Bi$_2$Se$_3$, 20 QLs Bi$_2$Te$_3$ and 1 QL Bi$_2$Se$_3$ / 19 QLs Bi$_2$Te$_3$ at different temperatures. (j) and (k) Resistance vs tilt angle of 20 QLs Bi$_2$Se$_3$ and 20 QLs Bi$_2$Te$_3$ in a fixed magnetic field of 160 kOe at $T$ = 2 K. The inset shows the rotation configuration. The magnetic field is always perpendicular to the current when the sample is rotating. (l) Resistance vs tilt angle of 1 QL Bi$_2$Se$_3$ / 19 QLs Bi$_2$Te$_3$ in a fixed magnetic field of 150 kOe at $T$ = 2 K. The function $|\cos\theta|$ fitting is shown as solid red lines in (j) – (l).

**Figure 4. WAL effect studied on 20 QLs Bi$_2$Se$_3$, 20 QLs Bi$_2$Te$_3$ and 1 QL Bi$_2$Se$_3$ / 19 QLs Bi$_2$Te$_3$ films.** (a) - (c) Normalized MR of 20 QLs Bi$_2$Se$_3$, 20 QLs Bi$_2$Te$_3$ and 1 QL Bi$_2$Se$_3$ / 19 QLs Bi$_2$Te$_3$ in the low field at different temperatures. (d) - (f) Fitted curves (solid red lines) with Eq. (1) are shown in 20 QLs Bi$_2$Se$_3$, 20 QLs Bi$_2$Te$_3$ and 1 QL Bi$_2$Se$_3$ / 19 QLs Bi$_2$Te$_3$ in perpendicular field at $T$ = 4 K respectively. For the 20 QLs Bi$_2$Se$_3$ film and the 1 QL Bi$_2$Se$_3$ / 19 QLs Bi$_2$Te$_3$ film, the fitting range is from 2.5kOe to -2.5kOe. However, for the 20 QLs Bi$_2$Te$_3$ film, the WAL dip is small that the fitting range is from 1kOe to -1kOe. (g) - (i) MR as a function of the normal H component with the $\theta = 90$º MR subtracted for 20 QLs Bi$_2$Se$_3$, 20 QLs Bi$_2$Te$_3$ and 1 QL Bi$_2$Se$_3$ / 19 QLs Bi$_2$Se$_3$ at $\theta$ equals to 0º, 45º, 60º, 80º. The inset shows the rotation configuration. (j) - (l) Violet symbols are the $T$ = 4 K magneto conductance for the 20 QLs Bi$_2$Se$_3$ film (j), the 20 QLs Bi$_2$Te$_3$ film (k) and the 1 QL Bi$_2$Se$_3$ / 19 QLs Bi$_2$Te$_3$ film (l) after subtracting the 3D WAL effect. The solid red lines show the fitting results with Eq. (1). Compared with corresponding figure (d) to (f), all the phase coherence lengths of our TI films become larger after subtracting the 3D WAL effect.

Figure 1

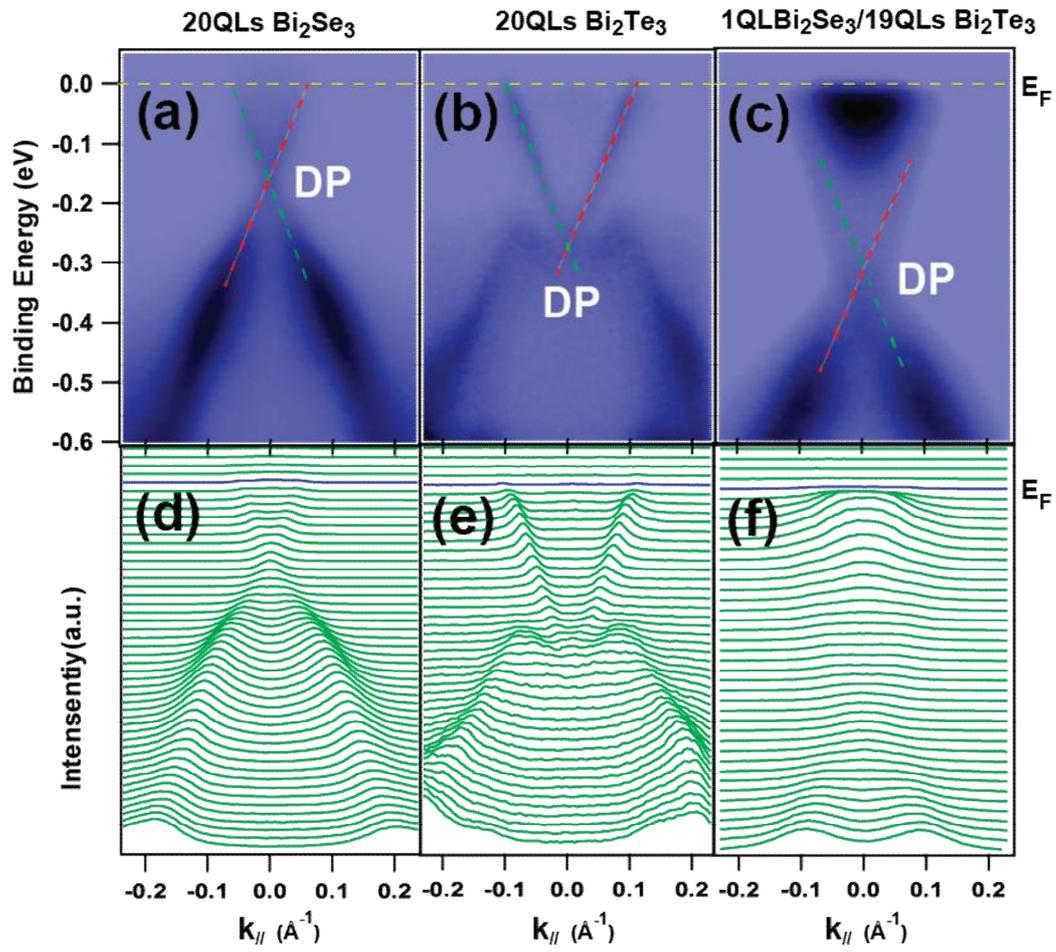

**Figure 2**

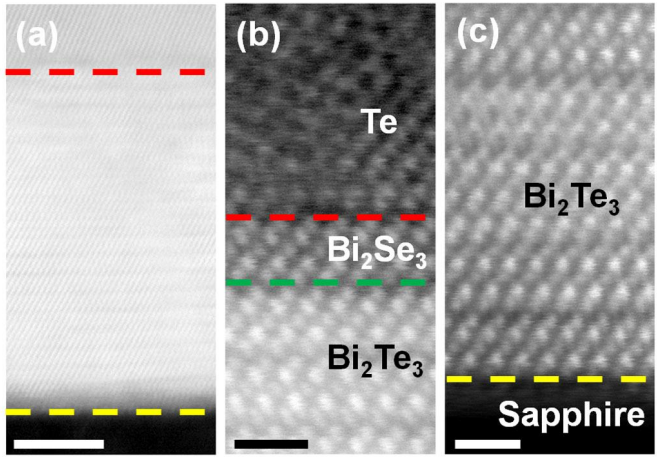

**Figure 3**

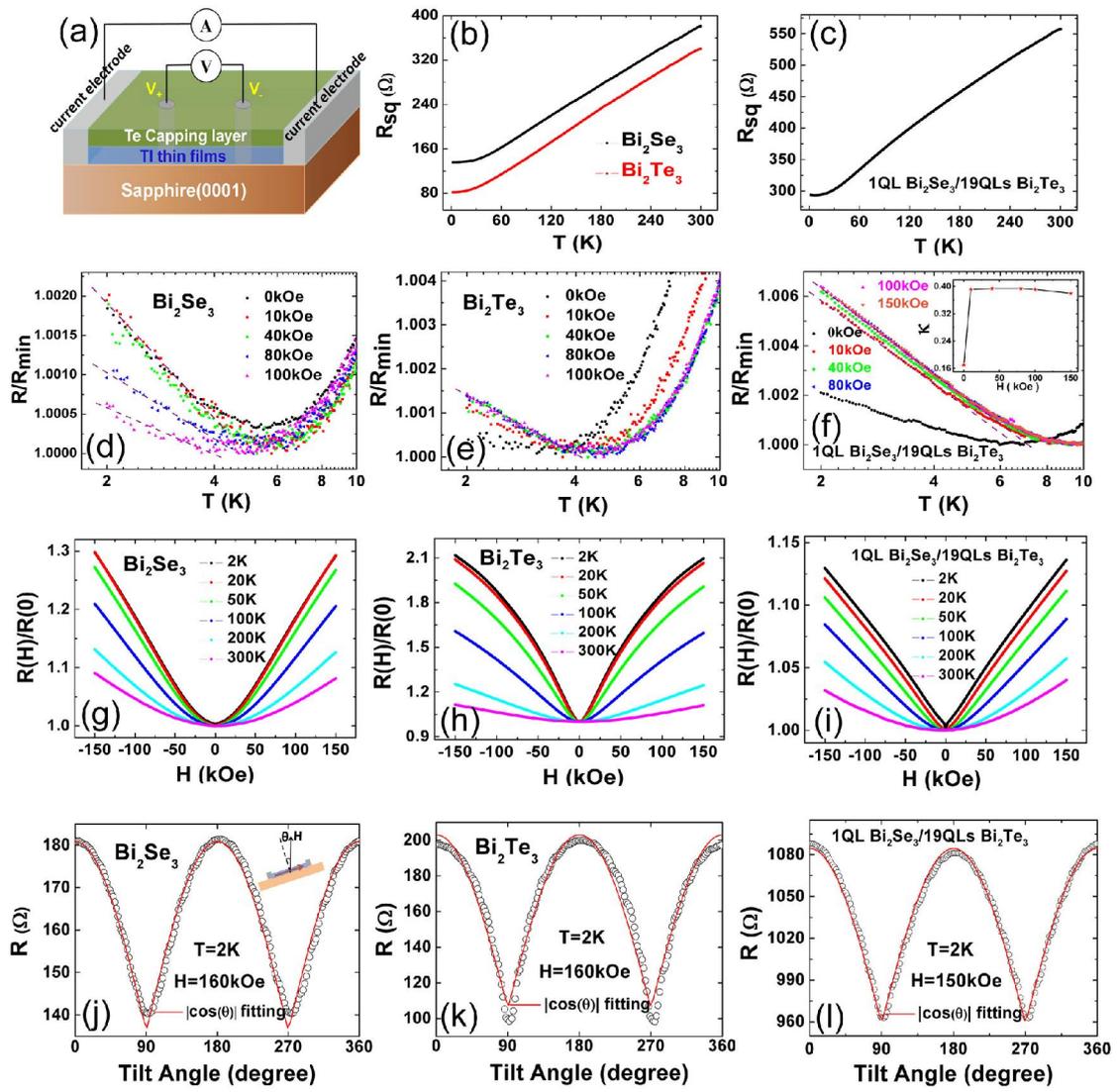

**Figure 4**

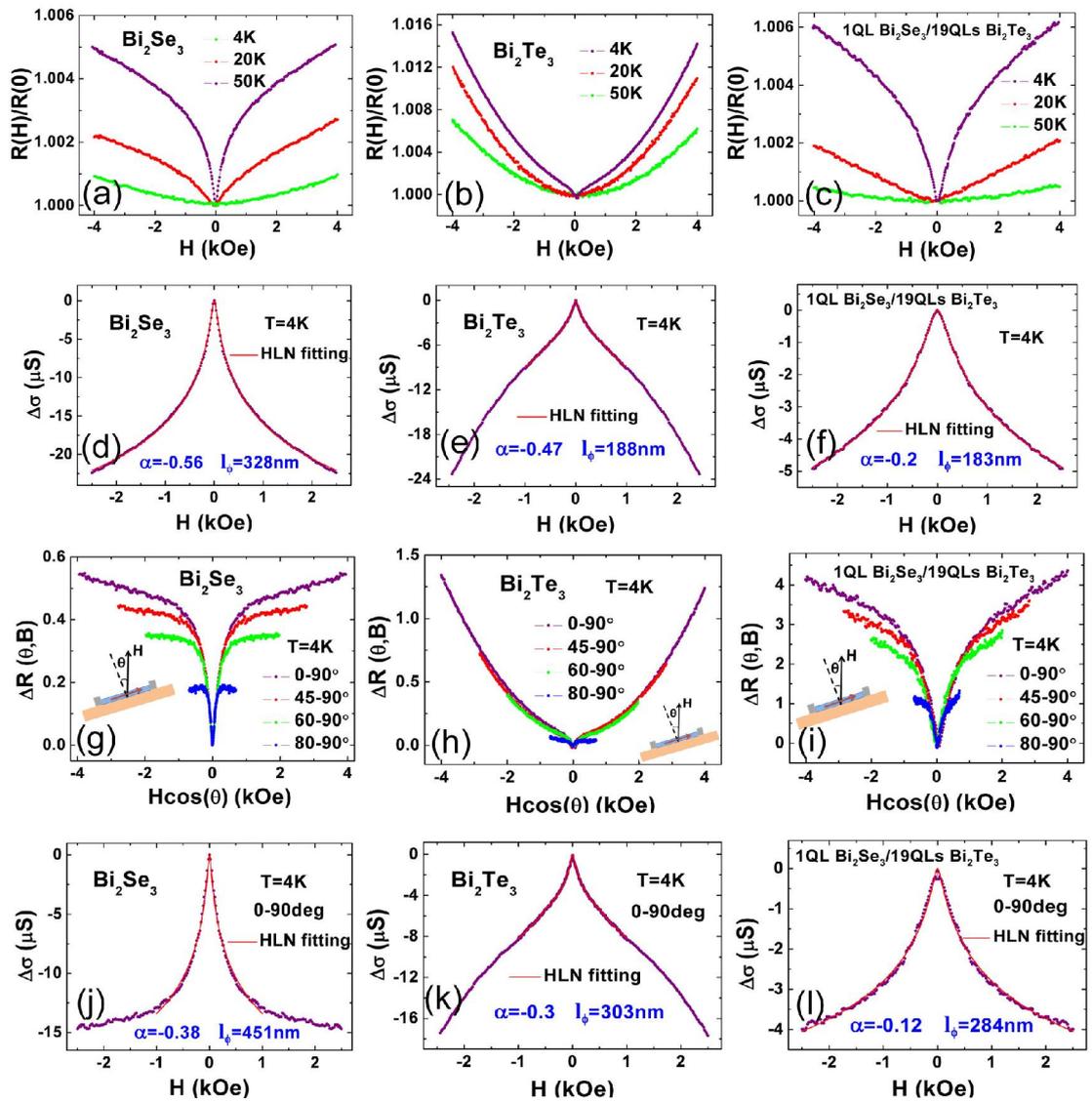

**Supplementary Information**

**Demonstration of surface transport in a hybrid $Bi_2Se_3$/$Bi_2Te_3$ heterostructure**

Yanfei Zhao[1, #], Cui-Zu Chang[2, 3#], Ying Jiang[4], Ashley DaSilva[5], Yi Sun[1], Huichao Wang[1], Ying Xing[1], Yong Wang[4], Ke He[2,*], Xucun Ma[2], Qi-Kun Xue[2, 3], and Jian Wang[1, *]

[1]International Center for Quantum Materials, School of Physics, Peking University, Beijing 100871, People's Republic of China

[2]Institute of Physics, Chinese Academy of Sciences, Beijing 100190, China

[3]State Key Lab of Low-Dimensional Quantum Physics, Department of Physics, Tsinghua University, Beijing 100084, China

[4]Center of Electron Microscopy, State Key Laboratory of Silicon Materials, Department of Materials Science and Engineering, Zhejiang University, Hangzhou, 310027, China

[5]Department of Physics, University of Texas at Austin, Austin, Texas 78712-1081, USA

[#]Authors equally contributed to this work.

*Correspondence and requests for materials should be addressed to J. W.

(jianwangphysics@pku.edu.cn) and K. H. (kehe@iphy.ac.cn).

# Supplementary Information

## Hall Measurements of 20 QLs $Bi_2Se_3$, 20 QLs $Bi_2Te_3$ and 1 QL $Bi_2Se_3$ / 19 QLs $Bi_2Te_3$

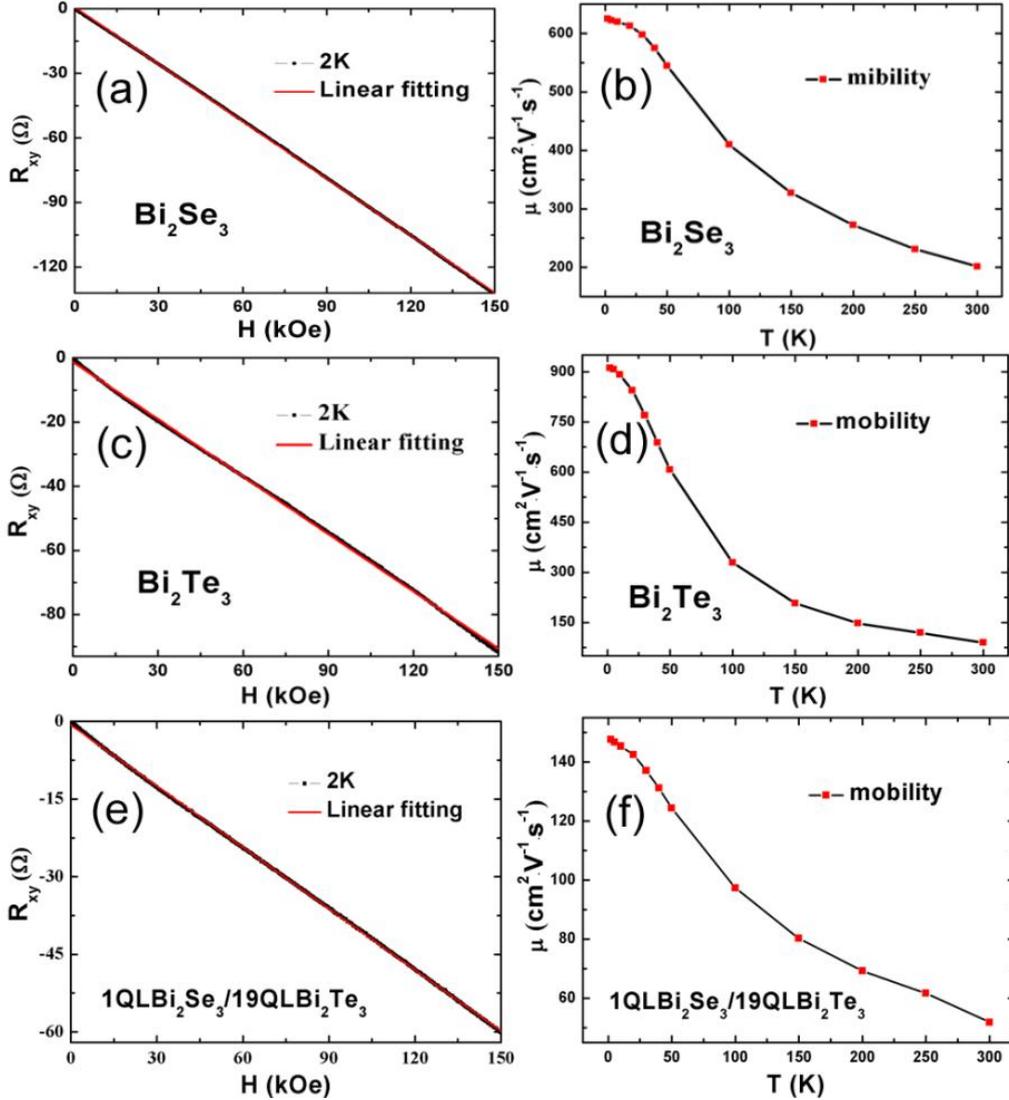

Fig. S1: Hall resistance of the 20 QLs $Bi_2Se_3$ film (a), 20 QLs $Bi_2Te_3$ film (c) and 1 QL $Bi_2Se_3$ / 19 QLs $Bi_2Te_3$ thin film (e) at 2 K. Temperature dependence of mobility for three samples, (b) for 20 QLs $Bi_2Se_3$, (d) for 20 QLs $Bi_2Te_3$ and (f) for 1 QL $Bi_2Se_3$ / 19 QLs $Bi_2Te_3$ thin films.

We determine the carrier density in the three samples using the Hall effect, as shown in Fig. S1. The carriers are electrons in all three samples. From the approximately linear behavior of $R_{xy}$ at low magnetic field, we estimate a carrier density of $3.53\times10^{19} cm^{-3}$ and mobility of 650$cm^2$/vs for the 20 QLs $Bi_2Se_3$ film at 2 K. For the 20QLs $Bi_2Te_3$ film, carrier density is $4.17\times10^{19} cm^{-3}$ and mobility is 912$cm^2$/vs. The carrier density of the 1 QL $Bi_2Se_3$ / 19 QLs $Bi_2Te_3$ film is $7.9\times10^{19} cm^{-3}$ and the mobility is 134$cm^2$/vs.

**Complete transport data of Fig. 3f**

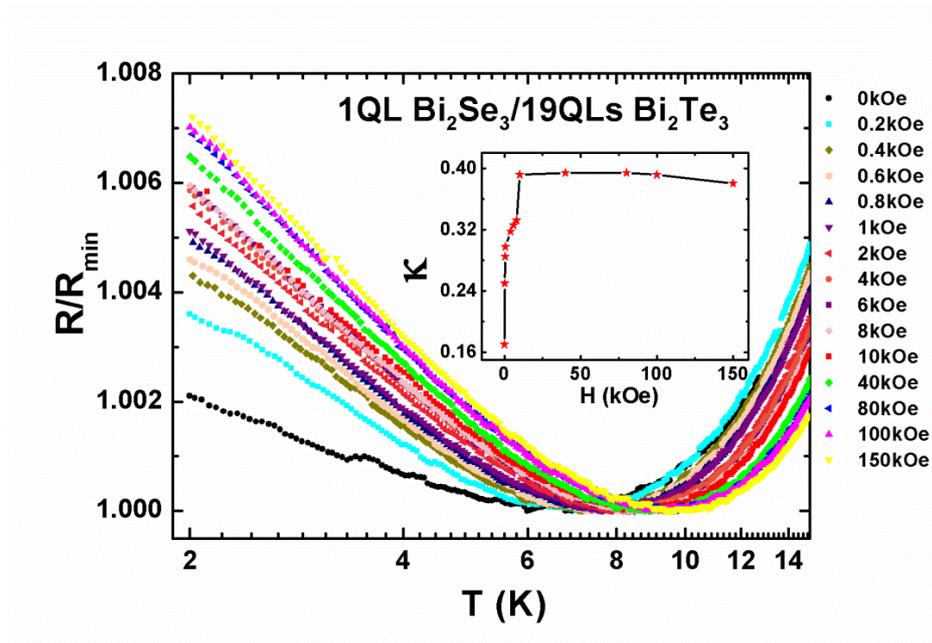

Fig. S2: The complete transport data on the 1 QL $Bi_2Se_3$ / 19 QLs $Bi_2Te_3$ film as a function of temperature at different magnetic field, indicates that a logarithmic increase with decreasing T. In the upper inset, the slope defined as $\kappa = (\pi h/e^2) d\sigma_{xx}/d\ln T$ is shown as a function of magnetic field. The $\kappa$ increases with increasing the magnetic field and becomes a constant when the magnetic field ranges from 10 kOe to 150 kOe.

## The Linear MR of the 1 QL $Bi_2Se_3$ / 19 QLs $Bi_2Te_3$ film at different tilt angle

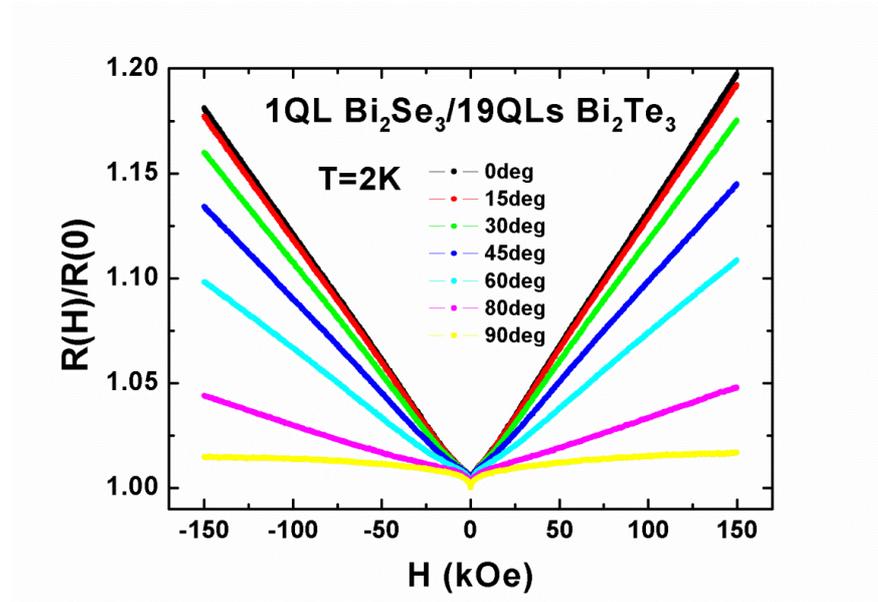

Fig. S3: Normalized MR vs magnetic field of the 1 QL $Bi_2Se_3$ / 19 QLs $Bi_2Te_3$ film at 2 K at different tilt angle between sample surface and the magnetic field.

In perpendicular magnetic field (θ=0deg), the sample exhibited a linear, non-saturated behavior across a wide range of magnetic field from 10 kOe to 150 kOe. The behavior gradually decreased when the sample was tilted away from the perpendicular angle and eventually became nonlinear in the parallel magnetic field (θ=90deg). It indicates that the linear MR is a 2D magneto-transport behavior.

**The fitting combined weak antilocalization with e-e interaction theory of 1 QL Bi$_2$Se$_3$ / 19 QLs Bi$_2$Te$_3$ film in low field at *T*=4K**

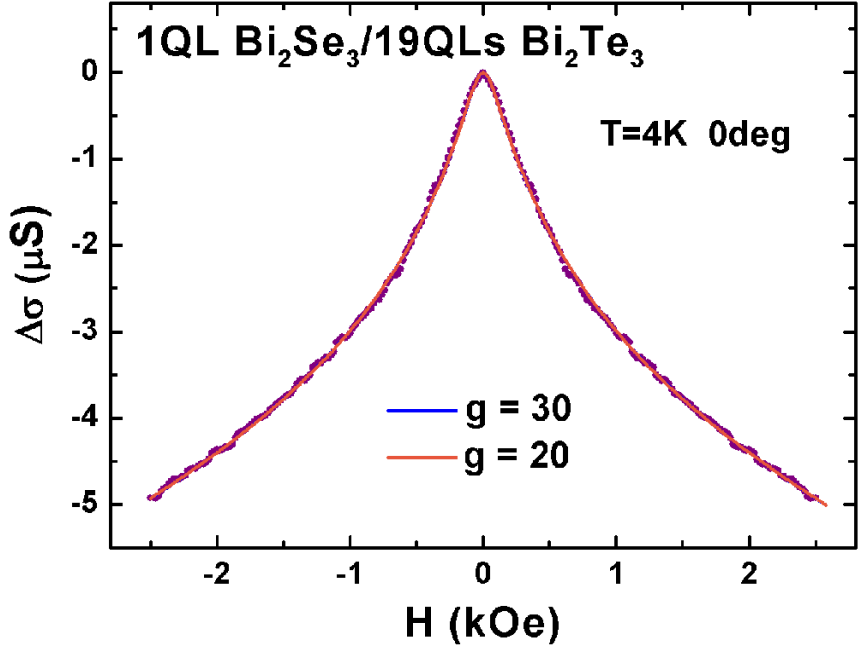

Fig. S4: Solid lines are the results of a combined WAL and EEI theory[S1]. Considering the g-factor of 1 QL Bi$_2$Se$_3$ / 19 QLs Bi$_2$Te$_3$ film is not so sure, we fixed g-factor to be 30 and 20. The fitting curves are plotted by blue and orange respectively.

The combined model [S1]: $\Delta\sigma(T, H_\perp) = \Delta\sigma_{WAL}(T, H_\perp) + \Delta\sigma_{EEI}(T, H_\perp)$

$$\Delta\sigma_{WAL} = -\frac{\alpha e^2}{2\pi^2 \hbar}\left[\ln\left(\frac{\hbar}{4Bel_\phi^2}\right) - \psi\left(\frac{1}{2} + \frac{\hbar}{4Bel_\phi^2}\right)\right] \text{ and } \Delta\sigma_{EEI} = -\frac{e^2}{4\pi^2\hbar}\tilde{F}_\sigma g_2(T,H)$$

Where $g_2(T,H) = \int_0^\infty d\Omega \ln\left|1 - \left(\frac{g\mu_B H/k_B T}{\Omega}\right)^2\right| \frac{d^2}{d\Omega^2}\frac{\Omega}{e^\Omega - 1}$

$\alpha$, $\tilde{F}_\sigma$ and $l_\phi$ are taken as fitting parameters.

The fitting parameters are as follows:

| g-factor | $\alpha$ | $l_\phi$ | $\tilde{F}_\sigma$ |
|---|---|---|---|
| 30 | -0.32 | 194nm | 0.6 |
| 20 | -0.25 | 193nm | 0.93 |

Reference
S1. Wang, J. et al. Evidence for electron-electron interaction in topological insulator thin films. *Phys. Rev. B.* **83**, 245438 (2011).